\documentclass[aps,prd,preprintnumbers,nofootinbib,showpacs,superscriptaddress]{revtex4}
\usepackage[dvipdfmx]{graphicx}
\usepackage{bm,latexsym,amsmath,amssymb,amsfonts} 
\DeclareMathAlphabet{\mathpzc}{OT1}{pzc}{m}{it}
\usepackage{color}
\usepackage[normalem]{ulem}
\usepackage{hyperref}

\begin{document}
\title{Relativistic quantum bouncing particles in a homogeneous gravitational field} 

\author{Ar Rohim}
\affiliation{Department of Physical Science, Graduate School of Science, Hiroshima University,
  Higashi-Hiroshima, 739-8526, Japan}
\author{Kazushige Ueda}
\author{Kazuhiro Yamamoto}
\affiliation{Department of Physics, Kyushu University, 
  Fukuoka,  819-0395, Japan}
\author{Shih-Yuin Lin}
\affiliation{Department of Physics, National Changhua University of Education,
  Changhua 50007, Taiwan}

\begin{abstract}
In this paper, we study the relativistic effect on the wave functions for a bouncing particle in a gravitational field. Motivated by the equivalence principle,	we investigate the Klein-Gordon and Dirac equations in Rindler coordinates with the boundary conditions mimicking a uniformly accelerated mirror in Minkowski space. In the nonrelativistic limit, all these models in the comoving frame reduce to the familiar eigenvalue problem for the Schr\"odinger equation with a fixed floor in a linear gravitational potential, as expected. We find that  
the transition frequency between two energy levels of a bouncing Dirac particle is greater than the counterpart of a Klein-Gordon particle, while both are greater than their nonrelativistic limit. The different corrections to eigen-energies of particles of different nature are associated with the different behaviors of their wave functions around the mirror boundary. 

\end{abstract}


\maketitle
\section{Introduction}
Quantum effects under the influence of gravitational field are the starting points of understanding the general features of the systems that both relativity and quantum mechanics come into play.  One of such systems which can be tested in laboratory is the quantum bouncer, which is a bouncing particle trapped in a linear gravitational potential above a floor, so that the stationary bound states of the particle have discrete energy levels associated with the normalizable wave functions. 
While the quantum bouncer problem has been explored in many theoretical works (e.g.,~\cite{Langhoff71,Gibbs75,Onofrio,Gea,Rosu}) and has even been a standard example or exercise in textbooks of quantum mechanics (e.g.,~\cite{Landau,Sakurai,Schwinger}), the discrete spectrum of the gravitational bound states were not probed experimentally until this century \cite{Nesvizhevsky,Nesvizhevsky1,Nesvizhevsky2, Westphal, Ichikawa, Kamiya, Nez} using the ultracold neutrons (UCN). It is confirmed that the wave functions of neutrons in the waveguide with and without gravity are given by the Airy functions in a linear potential and the sine functions in an infinite square-well potential, respectively, as those described in textbooks. 

The UCN are the neutrons with kinetic energy lower than the step barrier of the Fermi pseudo-potential \cite{Fermi36} which represents the coherent strong interaction of the neutrons with the atomic nuclei in a medium. The materials of high pseudo-potential energy such as beryllium and optical glass can thus serve as the UCN mirrors, by which the UCN can be totally reflected at arbitrary incidence angles. With the kinectic energy less than about 300 neV and the de Broglie wavelength greater than about 50 nm, the UCN reflections can be well described in the context of nonrelativistic quantum mechanics with a step potential barrier of the macroscopic medium \cite{Kowalski93, AIFrank14}.

The UCN are first observed in Dubna \cite{Shapiro69, Po18} and Munich \cite{Steyerl69}. Soon after those observations, Ilya Frank developed the idea of the neutron optics \cite{IMFrank, AIFrank91}. The above mentioned quantum bouncer experiment is an application of the neutron optics in noninertial frame: it is about the reflection of the UCN by a boundary of optical glass at rest in a linear gravitational potential. There are similar neutron optics with the boundary conditions and/or the UCN in noninertial motion. For example, the transmission of the UCN through an accelerated silicon slab, analogous to the photon transmission through an accelerated dielectric slab \cite{Tanaka82}, has been investigated in Refs. \cite{AIFrank, AIFrank2a,AIFrank2,AIFrank3,AIFrank4}. 
The UCN forced to go around a cylindrical UCN mirror is observed to be in a centrifugal quantum state in Ref. \cite{Voronin08}, where the multiple reflections of the UCN by the cylindrical mirror can be described as bouncing in an approximately linear radial potential near the mirror \footnote{In Ref.~\cite{Voronin08}, the authors solved the quantum mechanical wave functions of the UCN in cylindrical coordinates, which is actually a reference frame for an inertial observer at rest at the origin. While the co-moving frame with a UCN is a noninertial frame where the UCN experiences a centrifugal force, we do not say the calculation in Ref.~\cite{Voronin08} is a ``use of the equivalence principle", anyway, since the inertial force here (centrifugal force) is not exactly what the equivalence principle in general relativity is referring to.}.

In this paper, we revisit the above quantum bouncer problem in the viewpoint of the equivalence principle of relativity \cite{QEP}: a free particle repeatedly caught and bounced by a uniformly accelerated floor in Minkowski space can be regarded as a particle in a linear gravitational potential bouncing on the floor at rest in Rindler coordinates. Analysis on the Klein-Gordon (KG) particles bounced by a uniformly accelerated perfect mirror has been done in Ref.~\cite{Saa}. Later, the study was generalized to the case of Dirac particles in Ref.~\cite{Boulanger}. Along the same line, we solve the KG and Dirac equations for free particles with the boundary condition mimicking a mirror situated at the origin of the Rindler coordinate system to investigate how the linear gravitational potential affects the structure of the bound states and energy levels of a relativistic bouncing particle, and then compare the results in the nonrelativistic limit with those in quantum mechanics. For the KG equation, since the lowest energy eigenstates approximately see an infinite Fermi pseudopotential barrier, we follow Ref.~\cite{Saa} to simply introduce a Dirichlet boundary condition of vanishing field amplitude right at the mirror surface (cf. \cite{Alberto3}). Then we recover the well known corrections to the energy levels straightforwardly. For the Dirac equation in Rindler coordinates, however, Dirichlet boundary conditions do not work, because the condition of no fermionic particles at the floor will force the Dirac wave functions vanish everywhere in space, which is totally trivial. This is why the problem of bound states for Dirac particles in Rindler coordinates considered earlier in Ref.~\cite{Boulanger} (cf. \cite{Nicolaevici, Alberto,Alberto2}) has to be solved by using alternative boundary conditions. In this paper, we follow Ref.~\cite{Nicolaevici}  to adopt the boundary condition given in the MIT bag model (BC-MIT) \cite{Chodos1,Chodos2}, which implies that the normal probability current and the scalar densities vanish at the floor \cite{Hosaka,Tsushima}.  This enables us to look in details into the quantum bouncers of Dirac particles such as neutrons, and then compare their energy levels and transition frequencies with the nonrelativistic particles.

In this paper, the BC-MIT simply serves as an approximation of the effective Fermi pseudo-potential barrier of the mirror. In our mirror model for the UCN, the boundary thickness is of the order of 0.1 nm, which is much greater than the size of a hadron. We expect that the BC-MIT would be sufficient at this atomic scale, though in nuclear physics the chiral (or little/cloudy) bag model \cite{CT75, Rho1, Rho2, Rho3, Theberge, Theberge2, Thomas, Hosaka, Tsushima} is more general and realistic than the MIT bag model for hadrons.

If the neutron-antineutron oscillation exists, it will effectively make neutron a Majorana fermion \cite{Mohapatra09, Mohapatra, Gardner,Baldo,Babu2, Gardner16, Gardner18, Berezhiani}. Thus, for completeness, we also consider the case of Majorana particles in this paper. We construct Majorana wave functions from Dirac wave functions with arbitrary spin orientations.  The same problem will arise when imposing Dirichlet boundary conditions, anyway, and we again introduce the BC-MIT. 

Similar quantum bouncer experiments using positronium (Ps) atoms in Rydberg state instead of the UCN have been proposed in Refs.~\cite{Crivelli2015, Dufour2015} (also see Refs.~\cite{Mills2002, Cassidy2014} on the proposed gravitational free-fall experiments of Ps atoms.) In those proposals, the Ps atoms take the advantage of their lighter mass, which result in denser energy levels in the nonrelativistic limit but larger relativistic corrections to transition frequency. We also examine whether these relativistic corrections would be significant in those proposed experiments.

This paper is organized as follows. In Sec. \ref{NRQM}, we briefly review the dynamics of a bouncing particle in nonrelativistic quantum mechanics. Then in Sec. \ref{secKGparticles}, we study the bound states of a massive KG particle in Rindler coordinates. In Sec. \ref{FermionRindler}, we consider massive Dirac and Majorana particles of arbitrary spin orientations in Rindler coordinates.
The energy levels as well as the probability, scalar, and normal current densities for these relativistic particles are compared with
the conventional results in nonrelativistic quantum mechanics in Sec. \ref{compareBP}. Finally, Sec. \ref{SummaryConclusion} is our summary and conclusion. For comparison, we calculate the relativistic correction to the Hamiltonian of a spin-0 or $1/2$ particle in a perturbative approach in Appendix \ref{perturbation}. Throughout this paper, we use unit $c=\hbar=1$.

\section{Bouncing particles in nonrelativistic quantum mechanics}
\label{NRQM}

Consider a quantum mechanical particle in a linear gravitational potential and bouncing above the floor at $z=0$ (see e.g., 
\cite{Landau, Sakurai, Schwinger, Onofrio, Gea, Kamiya, Rosu}). 
For simplicity, we are working in a (1+1)D spacetime and modeling the effect of the floor and the gravitational field by the potential 
\begin{equation}
  V(z)=\left\{
\begin{array}{ll}
 maz, ~~ &\text{for $z>0$},\\
  \infty, ~~ &\text{for $z\leq 0$},
\end{array}
\right.
\label{Potential}
\end{equation}
where $m$ is the mass of the particle and $a=g$ is the uniform gravitational acceleration experienced by the particle.
In this ideal case with an infinite barrier as the floor \footnote{In realistic cases the effective potential barrier of the floor or mirror is finite and may depend on the momentum of the bouncing particle \cite{Squires78, Kawai99}.}, 
the particle is totally restricted in the region $z > 0$, where the dynamics of the particle is described by the time-dependent Schr\"{o}dinger equation
\begin{eqnarray}
  i\frac{\partial}{\partial t}\Psi(z,t) = -\frac{1}{2m}\frac{\partial^2 \Psi(z,t)}{\partial z^2}+maz\Psi(z,t).
\label{timedependentsch}
\end{eqnarray}
The wave function $\Psi(z,t)$ of the particle satisfies the boundary conditions $\Psi(z=0,t)=0$ and $\Psi(z\to\infty,t)=0$, which make the wave function in a normalizable bound state with a discrete energy spectrum, as we will see below. 

\begin{figure}[t]
  \begin{center}
    \includegraphics[width=9cm]{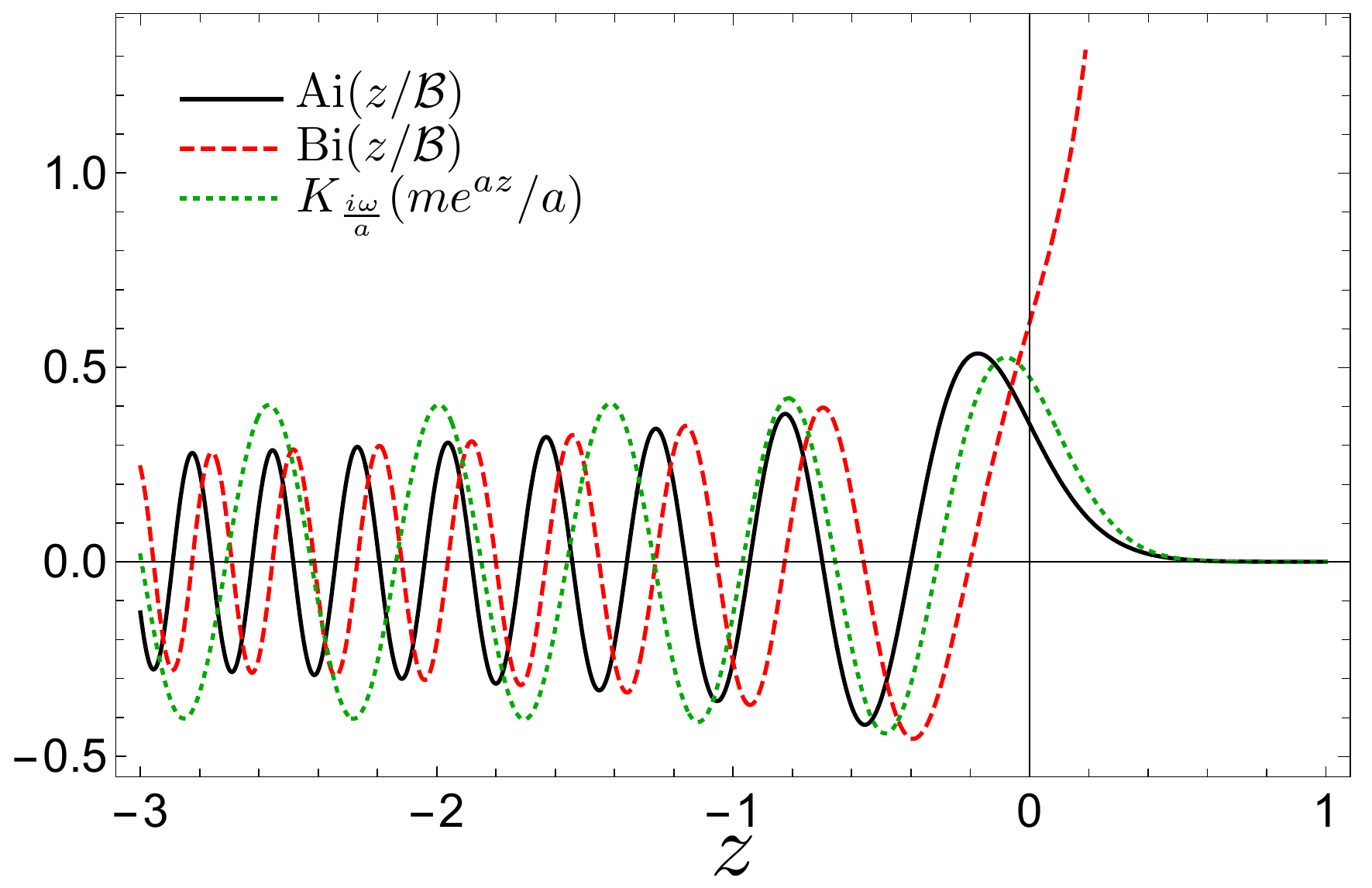}
    \caption{\label{AiryAiBiandBesselK}    Airy function Ai$(z/{\mathcal B})$, Airy function Bi$(z/{\mathcal B})$,
		and the (scaled) modified Bessel function of the second kind 
		$K_{\frac{i\omega}{a}}(m e^{a z}/a)$. Both Airy Ai and modified Bessel  $K$ converge to zero while Airy Bi  diverges to infinity as $z\to\infty$.
		Here $a=1$, $m=10$, $\omega=11$, and $\mathcal{B}$ is defined in (\ref{beta}).}
  \end{center}
\end{figure}

Suppose the stationary solution for Eq.~(\ref{timedependentsch}) has the form 
\begin{eqnarray}
\Psi(z,t)= e^{-iE_n t}\psi_n(z)
\end{eqnarray}
with the constant $E_n$ labeled by some index $n$ for later use. 
Then Eq.~(\ref{timedependentsch}) yields
\begin{eqnarray}
-\frac{1}{2m}\frac{d^2\psi_n(z)}{d z^2}+maz\psi_n(z)=E_n\psi_n(z).
\label{timeindependentsch}
\end{eqnarray} 
Introducing the dimensionless variable $\zeta=z/\mathcal{B}$  with the length scale 
\begin{eqnarray}
  \mathcal{B} = \left(2m^2 a\right)^{-1/3} ,
  \label{beta}
\end{eqnarray}
Equation~(\ref{timeindependentsch})
can be rewritten in the form
\begin{eqnarray}
-\frac{d^2\psi_n(\zeta)}{d\zeta^2}+(\zeta-\zeta_n)\psi_n(\zeta)=0,
\label{schbouncingmotion}
\end{eqnarray}
where $\zeta_n$ is defined by 
\begin{equation}
  E_n = ma\mathcal{B} \zeta_n. \label{zetanEn}
\end{equation}
Equation~(\ref{schbouncingmotion}) has two linearly independent solutions ${\rm Ai}(\zeta-\zeta_n)$ and ${\rm Bi}(\zeta-\zeta_n)$, 
which are the Airy functions (see FIG. \ref{AiryAiBiandBesselK}). Since Bi$(\zeta-\zeta_n$) diverges to infinity as $\zeta\to\infty$,
it will never satisfy the boundary condition $\Psi(z\to\infty,t)=0$, and so the bound-state wave function reads 
\begin{eqnarray}
\psi_n(\zeta) = \mathcal{N}_n\text{Ai}(\zeta-\zeta_n), \label{NRwavefunc}
\end{eqnarray}
where $-\zeta_n$ must be the $n$th zero of the Airy function such that Ai$(-\zeta_n)=0$, 
$n=1,2, \cdots$, to satisfy the mirror boundary condition of the wave function 
\begin{equation}
 0 = \Psi(z=0,t) = 
\mathcal{N}_n\text{Ai}(-\zeta_n)e^{-iE_n t},
\label{energylevel}
\end{equation}
for all $t$, while the normalization factor
\begin{equation}
  \mathcal{N}_n \equiv \left[\sqrt{\mathcal{B}}\, {\rm Ai}'\left(-\zeta_n \right)\right]^{-1}
	\label{normalizationfactor}
\end{equation}
is defined by $1=\int_{0}^{\infty}|\Psi_n(z)|^2dz
=\mathcal{N}_n^2\mathcal{B}\int_{0}^{\infty}\left[ {\rm Ai}\left(\zeta-\zeta_n\right) \right]^2 d\zeta
=\mathcal{N}_n^2\mathcal{B} \left[ {\rm Ai}'\left(-\zeta_n\right)\right]^2$.
Thus the eigen-energy $E_n$ in (\ref{zetanEn}) associated with the eigen-state $\psi_n$ in (\ref{NRwavefunc})
for some specific $n$ is proportional to the $n$th zero of Airy function $-\zeta_n$, 
whose negativity implies the positivity of the eigen-energy.

\section{Klein-Gordon bouncing particles}
\label{secKGparticles}

In what follows, we calculate the bound states of a spin-0 particle bouncing above a fixed floor in a linear gravitational potential. 
By the equivalence principle of relativity, the problem is equivalent to a free massive KG particle bounced  repeatedly by a uniformly accelerated floor. We will start with a review on the KG equation in Rindler coordinates \cite{Crispino} (cf. \cite{HIUY} for massless KG particles). Then, following Ref.~\cite{Saa}, we introduce a Dirichlet boundary condition at the floor to obtain the energy eigen-states of the KG bouncer in view of a Rindler observer co-moving with the floor. Results in this section may apply to the neutral, spinless composite bosons such as para-positronium (within its lifetime), and will be compared with those for the spin-1/2 bouncing particles later in Sec.~\ref{compareBP}.

\subsection{KG equation in Rindler coordinates}

The Rindler coordinate system is natural for a uniformly accelerated observer in Minkowski spacetime. 
Rindler coordinates $(\eta, x,y,\xi)$ can be transformed from Minkowski coordinates $(t,x,y,z)$ by 
\begin{eqnarray}
t=\frac{e^{a\xi}}{a}\sinh a\eta~~ {\rm and}~~ z=\frac{e^{a\xi}}{a}\cosh a\eta,
\end{eqnarray}
with $\eta, \xi \in {\bf R}$. Here the positive constant parameter $a$ is the proper acceleration of a point-like object uniformly accelerated in the $z$ direction and going along the worldline of $\xi=x=y=0$.
Then, the Rindler line element takes the form \cite{Lass63, BD82},
\begin{eqnarray}
ds^2&=&dt^2-dx^2-dy^2-dz^2\nonumber\\
&=& e^{2a\xi}\left(d\eta^2-d\xi^2\right)-d{\bm x}^2_\perp,
\label{Rindlermetric}
\end{eqnarray}
where ${\bm x}_\perp\equiv(x,y)$ coordinatize the two-dimensional subspace 
perpendicular to the direction of acceleration. 
Consider a free, massive scalar field $\phi$ in four-dimensional curved spacetime described by the action
\begin{eqnarray}
S=\frac{1}{2}\int d^4x \sqrt{-g}\left(g^{\mu\nu}\partial_\mu\phi\partial_\nu\phi-m^2\phi^2\right),
\label{KGaction}
\end{eqnarray}
where $m$ is the mass of field. Variation of the above action gives the KG equation 
\begin{eqnarray}
  \frac{1}{\sqrt{-g}}\partial_\mu\left(\sqrt{-g}g^{\mu\nu}\partial_\nu\phi\right)  +m^2\phi=0. 
\end{eqnarray} 
Inserting the metric in the second line of (\ref{Rindlermetric}), one obtains the KG equation 
\begin{eqnarray}
\left(\frac{\partial^2}{\partial \eta^2}-\frac{\partial^2}{\partial \xi^2}-
e^{2a\xi}\frac{\partial^2}{\partial {\bm x}^2_\perp}\right)\phi +m^2e^{2a\xi}\phi=0 \label{KGRindler}
\end{eqnarray}
in Rindler coordinates.
The normalized positive-energy solution for the above equation is well known (see, e.g., ~\cite{Crispino}):
\begin{eqnarray}
\phi\sim 
v^R_{\omega{\bm k}_\perp}
(\eta,\xi,{\bm x}_\perp)=\sqrt{\frac{\sinh(\pi\omega/ a)}{4\pi^4 a}}
K_{\frac{i\omega}{a}}\left(\frac{\kappa}{a}  e^{a\xi} \right)e^{i{\bm k}_\perp\cdot{\bm x}_\perp-i\omega\eta},
\label{KGmodeFree}
\end{eqnarray}
where $\omega\in (~0, \infty~)$, $\kappa \equiv \sqrt{m^2+{\bm k}_\perp^2}$ with ${\bm k}_\perp$ the component of momentum
perpendicular to the direction of acceleration, and $K_\nu(X)$ is the modified Bessel function of the second kind (FIG.~\ref{AiryAiBiandBesselK}).

\subsection{Bound states of a KG bouncing particle in Rindler coordinates}
\label{RelativCorrectionKGRindlerenergy}

Suppose the uniformly accelerated infinite floor is extended in the $xy$-plane and moving along the worldline of $\xi=0$ in the $tz$-subspace. So we introduce the Dirichlet boundary condition $\phi |_{\xi=0}=0$ to the positive-energy mode (\ref{KGmodeFree}), which implies \cite{Saa}
\begin{eqnarray}
  K_{\frac{i\omega}{a}}\left(\frac{\kappa}{a} \right)=0.
  \label{Kiom}
\end{eqnarray}
The above boundary condition requires that $\kappa/a$ must be one of the zeros of $K_{i\omega/a}(x)$,
and only a discrete set of the values of $\omega$ can satisfy Eq.~(\ref{Kiom}) with $\kappa/a$ fixed as has been shown by Refs.~\cite{Saa, Boulanger}. 
The energy levels of a free KG particle satisfying (\ref{Kiom}) in Rindler coordinates with 
${\bm k}_\perp=0$ (which implies $\kappa=m$) will be given in Section \ref{compareBP}.

\section{Dirac and Majorana bouncing particles} 
\label{FermionRindler}

Nexts, we turn to Dirac and Majorana particles of arbitrary spin orientations in Rindler coordinates.  
We follow the procedure in Refs. \cite{BD82, Crispino, Matsas, Vanzella, Suzuki, KUeda} 
to obtain the solutions for Dirac wave functions, but alternatively, in the Majorana representation.
The obtained solutions can immediately be transformed to the ones in the Dirac representation \cite{Aste, Pal}, 
and applied to construct the solutions for Majorana wave functions by taking the superpositions of 
positive- and negative-energy solutions. As we mentioned earlier, the boundary conditions for Dirac and Majorana particles are not trivial. In contrast to KG particles, imposing Dirichlet boundary conditions to Dirac and Majorana particles only yields trivial solutions. We will instead adopt the BC-MIT \cite{Chodos1,Chodos2} for our Dirac and Majorana particles.

\subsection{Dirac particles in Rindler coordinates} 
\label{DiracParti}

The Dirac equation for a Dirac wave function $\tilde{\psi}^{\rm D}$ in the Majorana representation in Rindler coordinates reads \cite{BD82}
\begin{equation}
\left[ i \tilde \gamma^\mu_{\rm R} \left(\frac{\partial}{\partial x^\mu}+\tilde\Gamma_\mu\right)-m \right] \tilde{\psi}^{\rm D}=0.
\label{DiracEqRindler}
\end{equation}
Here, $m$ is the rest mass of the particle, and $\tilde\Gamma_\mu$ is the spin connection given by
\begin{eqnarray}
\tilde\Gamma_\mu &=& 
\frac{1}{2}\Sigma^{AB} V_A^\nu \nabla_\mu V_{B\nu} = 
\frac{1}{2}\Sigma^{AB} V_A^\nu \left(\partial_\mu V_{B\nu} -\Gamma_{\mu\nu}^\rho V_{B\rho} \right) \nonumber\\
&=& \left(\frac{a}{2} \tilde{\gamma}^0 \tilde{\gamma}^3,~0,~0,~0\right),
\label{spinconnection}
\end{eqnarray}
where $\Sigma^{AB} \equiv \frac{1}{4} [\gamma^A, \gamma^B]$, the vierbien $V_A^\mu= {\rm diag}\{ e^{-a\xi},1,1,e^{-a\xi}\}$ is defined by $g^{\mu\nu}=\eta^{AB}V_A^\mu V_B^\nu$ with $\eta^{AB}={\rm diag}\{1,-1,-1,-1\}$, and  
$\tilde\gamma^\mu_{\rm R} \equiv \tilde\gamma^A V_A^\mu$ are the gamma matrices in the Majorana representations in Rindler coordinates, such that $\{ \tilde\gamma^A ,\tilde\gamma^B \}=2\eta^{AB}$ and $\{ \tilde{\gamma}^\mu_{\rm R} , \tilde{\gamma}^\nu_{\rm R} \}=2g^{\mu\nu}$.
Here and below, we use $\tilde{\psi}$ and $\psi$ to denote the wave functions in the Majorana and Dirac representations, respectively, and $\tilde{\gamma}^\mu$ and $\gamma^\mu$ to denote the gamma matrices in the Majorana and Dirac representations in Minkowski coordinates \cite{Pal}. 
The explicit forms of $\tilde\gamma^\mu_{\rm R}$ and $\tilde{\gamma}^\mu$ can be derived from $\gamma^\mu$ by using the unitary transformation \cite{Aste}
\begin{eqnarray}
\tilde{\gamma}^\mu=U{\gamma}^\mu U^\dagger,~~U=U^\dagger=U^{-1}=\frac{1}{\sqrt{2}}\begin{pmatrix}
{\bf 1} & {\bf\sigma}_2\\
\sigma_2 &-{\bf 1} 
\end{pmatrix},
\label{MDM}
\end{eqnarray}
with the Pauli matrix ${\bf \sigma}_2 = \left(\begin{array}{cc} 0 & -i \\ i & 0\end{array}\right) $ 
and the $2\times 2$ identity matrix ${\bf 1}$. 
In this paper, we use the gamma matrices $\gamma^\mu$ of Refs. \cite{Greiner00, IZ80} in the Dirac representation in Minkowski coordinates. From Eq.~(\ref{MDM}), the gamma matrices $\tilde\gamma^\mu$ in the Majorana representation in Minkowski coordinates satisfy $(\tilde\gamma^\mu)^*=-\tilde\gamma^\mu$. Explicitly, $\tilde\gamma^\mu$ are purely imaginary \cite{Aste, Pal}.

\subsubsection{Wave packets of positive energy}

A wave packet of a Dirac particle can be expanded as 
\begin{eqnarray}
\tilde{\psi}^{\rm D}(\eta,\xi,{\bm x_\perp})=\sum_{\sigma=\pm} \int_{-\infty} ^{\infty}\frac{d{\bm {\bm k_\perp}}}{2\pi} \int_0 ^{\infty} d\omega \left[ b(\omega, {\bm {\bm k_\perp}},\sigma)\tilde{\psi}_{\omega{\bm {\bm k_\perp}} \sigma}^{\rm D+} (\eta,\xi,{\bm x_\perp})+
d^*(\omega, {\bm {\bm k_\perp}},\sigma) \tilde{\psi}^{\rm D-}_{\omega{\bm {\bm k_\perp}} \sigma} (\eta,\xi,{\bm x_\perp})\right],
\label{Dirac}
\end{eqnarray}
where $b(\omega, {\bm {\bm k_\perp}},\sigma)$ and $d^*(\omega, {\bm {\bm k_\perp}},\sigma)$ are the amplitudes for the waves
of positive- and negative-energy, respectively. Let us consider the waves of positive-energy only and set $d^*(\omega,{\bm{\bm k_\perp}},\sigma)=0$ \cite{Greiner00}, which is a good approximation for cold neutrons ($|{\bm k}| \ll m$) \cite{IZ80}. 
We postulate an ansatz for the positive-energy solution as 
\begin{eqnarray}
\tilde{\psi}_{\omega {\bm {\bm k_\perp}}\sigma}^{\rm D+}\equiv \tilde{f}_{\omega{\bm {\bm k_\perp}} \sigma}^{\rm D}(\xi) e^{i{\bm {\bm k_\perp}}\cdot{\bm x_\perp}} e^{-i\omega \eta},
\label{ansatzpsi}
\end{eqnarray}
where  $\tilde{f}_{\omega{\bm {\bm k_\perp}} \sigma}^{\rm D} (\xi)$ denotes a Dirac spinor in the Majorana representation defined through the 
{\it two-component spinors} $\tilde{\chi}_1(\xi)$ and $\tilde{\chi}_2(\xi)$ as 
\begin{eqnarray}
\tilde{f}_{\omega {\bm {\bm k_\perp}}\sigma}^{\rm D} (\xi)=
\begin{pmatrix}
\tilde{\chi}_1 (\xi) \\
\tilde{\chi}_2 (\xi)
\label{Diracspinorcomponents}
\end{pmatrix}.
\end{eqnarray}
Then the Dirac equation reads 
\begin{eqnarray}
\frac{\omega}{a}\tilde{f}_{\omega{\bm {\bm k_\perp}} \sigma}^{\rm D} (\xi)
=\biggl[ \frac{m}{a} e^{a\xi} \tilde\beta - \frac{i}{2}\tilde\alpha_3 -\frac{i}{a} \tilde\alpha_3 
\frac{\partial}{\partial\xi}+\frac{k_1}{a}e^{a\xi} \tilde\alpha_1 +\frac{k_2}{a}e^{a\xi} \tilde\alpha_2 
\biggr]\tilde{f}^{\rm D}_{\omega{\bm {\bm k_\perp}} \sigma} (\xi) 
\label{dirac}
\end{eqnarray}
with $\tilde\beta \equiv \tilde{\gamma}^0$ and $\tilde\alpha_j \equiv \tilde\gamma^0\tilde\gamma^j$, or,
after inserting (\ref{Diracspinorcomponents}), 
\begin{eqnarray}
\omega \tilde{\chi}_1(\xi) = me^{a\xi} \sigma_2\tilde{\chi}_2(\xi)+i\frac{a}{2}\sigma_3\tilde{\chi}_2(\xi)+i\sigma_3 
\frac{\partial\tilde{\chi}_2(\xi)}{\partial \xi}-e^{a\xi} k_1\sigma_1\tilde{\chi}_2(\xi)+e^{a\xi} k_2 \tilde{\chi}_1(\xi),
\label{42}
\\
\omega\tilde{\chi}_2(\xi) = me^{a\xi} \sigma_2\tilde{\chi}_1(\xi)+i\frac{a}{2}\sigma_3\tilde{\chi}_1(\xi)+i\sigma_3 
\frac{\partial \tilde{\chi}_1(\xi)}{\partial \xi}-e^{a\xi} k_1\sigma_1\tilde{\chi}_1(\xi)-e^{a\xi} k_2\tilde{\chi}_2(\xi),
\label{43}
\label{2equation}
\end{eqnarray}
where $\sigma_1$ and $\sigma_3$ are the Pauli matrices, too. 
Further calculation shows 
\begin{eqnarray}
\frac{1}{a} \frac{\partial}{\partial\xi}\left(\frac{1}{a}\frac{\partial}{\partial\xi}\tilde{\chi}_1\right)=\left[\left(m^2+{\bm {\bm k_\perp}}^2\right)\frac{1}{a^2}e^{2a\xi}+\frac{1}{4}-\frac{\omega^2}{a^2}\right]\tilde{\chi}_1+ \frac{i\omega}{a}\sigma_3\tilde{\chi}_2,
\label{44} \\
\frac{1}{a}\frac{\partial}{\partial\xi}\left(\frac{1}{a}\frac{\partial}{\partial\xi}\tilde{\chi}_2\right)=\left[\left(m^2+{\bm {\bm k_\perp}}^2\right)\frac{1}{a^2}e^{2a\xi}+\frac{1}{4}-\frac{\omega^2}{a^2}\right]\tilde{\chi}_2+\frac{i\omega}{a}\sigma_3\tilde{\chi}_1,
\label{45}
\end{eqnarray}
where $\tilde{\chi}_1$ and $\tilde{\chi}_2$ are coupled. To proceed, we introduce another two-component spinor
\begin{eqnarray}
\phi^{\pm}= \tilde{\chi}_1\mp\tilde{\chi}_2=
\begin{pmatrix}
\tilde{\vartheta}^\pm (\xi) \\
\tilde{\varsigma}^\pm (\xi)
\end{pmatrix},
\label{spinorcomponents}
\end{eqnarray}
then Eqs.~(\ref{44}) and (\ref{45}) yield 
\begin{eqnarray}
\left(\frac{1}{a}\frac{\partial}{\partial\xi}\frac{1}{a}\frac{\partial}{\partial\xi}\right)
\tilde{\vartheta}^\pm (\xi)
&=&\left[(m^2+{\bm {\bm k_\perp}}^2)\frac{1}{a^2}e^{2a\xi}+\left(\frac{i\omega}{a}\mp\frac{1}{2}\right)^2\right]
\tilde{\vartheta}^\pm (\xi) ,
\label{Bessel1} \\
\left(\frac{1}{a}\frac{\partial}{\partial\xi}\frac{1}{a}\frac{\partial}{\partial\xi}\right)
\tilde{\varsigma}^\pm (\xi)
&=&\left[(m^2+{\bm {\bm k_\perp}}^2)\frac{1}{a^2}e^{2a\xi}+\biggl(\frac{i\omega}{a}\pm\frac{1}{2}\biggr)^2\right]
\tilde{\varsigma}^\pm (\xi). 
\label{Bessel2}
\end{eqnarray}
The general solution for Dirac wave function in the case of general momentum and arbitrary spin orientation can be found, for example, in Refs.~\cite{Suzuki, KUeda}. As has been known from those references, the solutions regular as $\xi\to\infty$ for 
Eqs.~(\ref{Bessel1}) and (\ref{Bessel2}) are 
respectively given by
(cf. Eqs.~(\ref{KGRindler})-(\ref{KGmodeFree}))
\begin{equation}
\tilde{\vartheta}^\pm(\xi)=A_\pm K_\mp(\xi), 
\hspace{1cm}
\tilde{\varsigma}^\pm(\xi)=B_\pm K_\pm(\xi), 
\end{equation}
where  $K_\pm (\xi) \equiv K_{\frac{i\omega}{a}\pm \frac{1}{2}}(\frac{\kappa}{a}e^{a\xi})$, $\kappa=\sqrt{m^2+\bm{k}_\perp^2}$, 
and $A_\pm$ and $B_\pm$ are complex coefficients with the information of spin orientation included. Thus, the solution for $\tilde{f}_{\omega{\bm {\bm k_\perp}}\sigma}^{\rm D}$ is 
\begin{eqnarray}
~~\tilde{f}_{\omega{\bm {\bm k_\perp}}\sigma}^{\rm D}=\frac{1}{2}\begin{pmatrix}
\tilde{\vartheta}^++\tilde{\vartheta}^-\\
\tilde{\varsigma}^++\tilde{\varsigma}^-\\
-\tilde{\vartheta}^++\tilde{\vartheta}^-\\
-\tilde{\varsigma}^++\tilde{\varsigma}^-
\end{pmatrix}
=\frac{1}{2}\begin{pmatrix}
A_+ K_-(\xi) +A_-K_+(\xi) \\ 
B_+K_+(\xi) +B_-K_-(\xi) \\
-A_+K_-(\xi) +A_-K_+(\xi) \\
-B_+K_+(\xi) +B_-K_-(\xi) 
\end{pmatrix}~.
\label{fD1}
\end{eqnarray}
Substituting it back into the Dirac equation (\ref{dirac}), we get four linear relations between the coefficients (cf. Ref.~\cite{KUeda} in the Dirac representation),
\begin{eqnarray}
&&i\kappa A_++k_2A_-+(im+k_1)B_+=0~, \hspace{1cm} (im-k_1)A_-+k_2B_++i\kappa B_-=0~, \nonumber\\
&&k_2A_+-i\kappa A_--(im+k_1)B_-=0~, \hspace{1cm} -(im-k_1)A_+-i\kappa B_++k_2B_-=0~,
\label{ABrelations} 
\end{eqnarray}
which imply that $B_\pm$ can be represented in $A_\pm$, and the solution (\ref{ansatzpsi}) with (\ref{fD1}) for Dirac particles in the Majorana representation can be rewritten as
\begin{eqnarray}
\tilde{\psi}_{\omega {\bm {\bm k_\perp}}\sigma}^{\rm D+}
=\mathcal{N}_{\omega{\bm {\bm k_\perp}}\sigma}^{\rm D}e^{i{\bm {\bm k_\perp}}\cdot{\bm x_\perp}}e^{-i\omega \eta}
\begin{pmatrix}
A_+ K_-(\xi) +A_-K_+(\xi)\\ 
-\big(\frac{i\kappa A_++k_2A_-}{im+k_1}\big)K_+(\xi)+\big(\frac{k_2A_+-i\kappa A_-}{im+k_1}\big)K_-(\xi)\\ 
-A_+K_-(\xi)+A_-K_+(\xi)\\
\big(\frac{i\kappa A_++k_2A_-}{im+k_1}\big)K_+(\xi)+\big(\frac{k_2A_+-i\kappa A_-}{im+k_1}\big)K_-(\xi)
\end{pmatrix},
\label{DiracInMajoranaRep}
\end{eqnarray}
and in the Dirac representation as
\begin{eqnarray}
\psi_{\omega{\bm {\bm k_\perp}} \sigma}^{\rm D+}
=\frac{\mathcal{N}_{\omega{\bm {\bm k_\perp}}\sigma}^{\rm D} e^{i{\bm {\bm k_\perp}}\cdot{\bm x_\perp}}e^{-i\omega \eta}}{im+k_1}
\begin{pmatrix}
\big[\kappa A_++(im+k_1-ik_2)A_-\big]K_+(\xi)
+\big[(im+k_1-ik_2)A_+-\kappa A_-\big]K_-(\xi)\\ 
\big[-i\kappa A_++(-m+ik_1-k_2)A_-\big]K_+(\xi) 
+\big[(m-ik_1+k_2)A_+-i\kappa A_-\big]K_-(\xi)\\ 
-\big[\kappa A_++(im+k_1-ik_2)A_-\big]K_+(\xi) 
+\big[(im+k_1-ik_2)A_+-\kappa A_-\big]K_-(\xi)\\ 
\big[-i\kappa A_++(-m+ik_1-k_2)A_-\big]K_+(\xi)
-\big[(m-ik_1+k_2)A_+-i\kappa A_-\big]K_-(\xi)
\label{Diracspinor4dim}
\end{pmatrix}
\end{eqnarray}
after a unitary transformation $\psi_{\omega{\bm {\bm k_\perp}}\sigma}^{\rm D+}=U\tilde{\psi}_{\omega{\bm {\bm k_\perp}}\sigma}^{\rm D+}$ with the unitary matrix $U$ given in Eq.~(\ref{MDM}) \cite{Aste, Pal}.
Here, $\mathcal{N}^{\rm D}_{\omega{\bm {\bm k_\perp}}\sigma}$ is the normalization constant determined depending on the boundary condition (see below).
\subsubsection{Boundary conditions of ideal mirrors for Dirac particles}

If one introduces the Dirichlet boundary condition $\tilde{\psi}_{\omega{\bm {\bm k_\perp}} \sigma}^{\rm D+}|_{\xi=0}=\psi_{\omega{\bm {\bm k_\perp}} \sigma}^{\rm D+}|_{\xi=0}=0$ similar to (\ref{energylevel}) and (\ref{Kiom}) for Schr\"odinger and KG particles, one will immediately get $A_+=A_-=0$ from (\ref{DiracInMajoranaRep}), corresponding to the trivial solution $\tilde{\psi}_{\omega{\bm {\bm k_\perp}} \sigma}^{\rm D+}=0$ for all $\xi$. To obtain nontrivial analytical solutions for Dirac bouncers, one simple resolution is to apply the BC-MIT mimicking a mirror situated at $\xi=0$ \cite{Nicolaevici}. 
The BC-MIT for Dirac wave function $\psi^{\rm D}$ is explicitly given as 
\cite{Chodos1,Chodos2}
\begin{eqnarray}
\left. i N_\mu\gamma^\mu_{{\rm R}}\psi^{\rm D}\right|_{\xi=0}=\left.\psi^{\rm D}\right|_{\xi=0},
	\label{MIT1} 
\end{eqnarray}
where $N_\mu$ is the unit four-normal to the boundary ($N_\mu N^\mu = -1$), 
pointing into the region where the Dirac bouncer can possibly be found (i.e. $\xi>0$).
In our setup, we have $N_\mu= (0,0,0,e^{a\xi})$ and the gamma matrices $\gamma^\mu_{\rm R}$ in Rindler coordinates in (\ref{MIT1}),
and so $N_\mu\gamma^\mu_{{\rm R}}=\gamma^3$. 
Inserting the Dirac wave function of a single positive-energy mode (\ref{Diracspinor4dim}) into (\ref{MIT1}), one gets 
\begin{equation}
\left(\begin{array}{lr}
i\kappa K_+(0)+\left(im+k_1-ik_2\right)K_-(0) \,\, & \,\,-\kappa  K_-(0) + \left(-m+ik_1+k_2\right)K_+(0)\\
\kappa K_+(0)+ \left(m-ik_1+k_2\right)K_-(0)\,\, & \,\, -i\kappa K_-(0) -(im+k_1+ik_2)K_+(0)
\end{array}\right)
\left(\begin{array}{c}
A_+\\A_-
\end{array}\right) =
\left(\begin{array}{c}
0 \\ 0
\end{array}\right),
\label{boundstatesDirac4} 
\end{equation}
where $K_\pm(0) = K_{\frac{i\omega}{a}\pm\frac{1}{2}}\left(\frac{\kappa}{a}\right)$. When $k_2\not=0$, for nontrivial $A_\pm$, (\ref{boundstatesDirac4}) implies that the determinant of its $2\times 2$ coefficient matrix has to vanish. This gives
\begin{equation}
  \left[K_+(0)\right]^2 + \left[K_-(0)\right]^2 + 2 \frac{m}{\kappa} K_+(0) K_-(0) = 0, 
	\label{Kgeneral}
\end{equation}
which leads to a discrete spectrum of $\omega$, like the KG bouncers under the Dirichlet boundary condition (\ref{Kiom}).  
With a specific solution of $\omega$ for (\ref{Kgeneral}), one may freely choose a complex value for one of $A_\pm$, then the value of the other will be given by (\ref{boundstatesDirac4}).

When $|k_2|$ is very small but $|k_1|$ is not, one has either $|A_+| \ll |A_-|$, or $|A_-| \ll |A_+|$, depending on the solution of $\omega$ for (\ref{Kgeneral}). When $k_2=0$ exactly but $k_1\not=0$, (\ref{boundstatesDirac4}) yields 
\begin{eqnarray} 
  \left[(m-i k_1) K_{\frac{i\omega}{a}+\frac{1}{2}}\left(\frac{\kappa}{a}\right)+ 
    \kappa K_{\frac{i\omega}{a}-\frac{1}{2}}\left(\frac{\kappa}{a}\right)\right] A_- &=& 0, \nonumber\\
  \left[(m-i k_1) K_{\frac{i\omega}{a}-\frac{1}{2}}\left(\frac{\kappa}{a}\right) + 
    \kappa K_{\frac{i\omega}{a}+\frac{1}{2}}\left(\frac{\kappa}{a}\right)\right] A_+ &=& 0.
\end{eqnarray}
If both $A_+$ and $A_-$ are nonzero, the above two conditions will never be simultaneously satisfied since the expressions in the above two square brackets do not both vanish for the same set of parameter values. The bound states of Dirac bouncers 
exist only when either $A_+=0$ but $A_-\not=0$, or $A_-=0$ but $A_+\not=0$. The former implies the condition 
$(m-i k_1) K_{\frac{i\omega}{a}+\frac{1}{2}}\left(\frac{\kappa}{a}\right)+ 
    \kappa K_{\frac{i\omega}{a}-\frac{1}{2}}\left(\frac{\kappa}{a}\right)=0$, the latter implies 
$(m -i k_1) K_{\frac{i\omega}{a}-\frac{1}{2}}\left(\frac{\kappa}{a}\right) + 
    \kappa K_{\frac{i\omega}{a}+\frac{1}{2}}\left(\frac{\kappa}{a}\right)=0$, and
the product of these two conditions is nothing but (\ref{Kgeneral}) with $k_2=0$ but $k_1\not=0$. It is clear that the cases of $k_2=0$ but $k_1\not=0$ are the limiting cases of $k_2\to 0$ but $|k_1|$ finite.

When ${\bm k_\perp}=0$, condition (\ref{boundstatesDirac4}) reduces to
\begin{eqnarray}
K_{\frac{i\omega}{a}+\frac{1}{2}} \left(\frac{m}{a}\right) + K_{\frac{i\omega}{a}-\frac{1}{2}} \left(\frac{m}{a}\right)=0,
\label{Kiomhalf}
\end{eqnarray}
if $A_+$ and $A_-$ are not both vanishing\footnote{
Our bound states (\ref{Kiomhalf}) coincide with the result in Eq. (37) of Ref.~\cite{Boulanger} for $\kappa=m$ and their chosen parameter 
$\sigma=+1$. 
Their bound states of $\sigma=-1$ can be recovered by using the chiral MIT boundary condition of chiral angle $\Theta=\pi$ \cite{Lutken1984}, or explicitly given as $\left. i N_\mu\gamma^\mu_{{\rm R}}\psi^{\rm D}\right|_{\xi=0}=-\left.\psi^{\rm D}\right|_{\xi=0}$, with the same definition of the unit-normal to the boundary $N_\mu$ in the BC-MIT (\ref{MIT1}).}. 
 Then (\ref{boundstatesDirac4}) does not introduce any relation between $A_+$ and $A_-$. 
If this special case is approached by letting $(k_1, k_2) \to (0,0)$, while a relation between $A_+$ and $A_-$ will be given by (\ref{boundstatesDirac4}) as $(k_1, k_2) \not= (0,0)$, the relation will depend on how $(k_1, k_2)$ approaches $(0,0)$. 
Every choice of $A_\pm$ can be connected to some way of $(k_1, k_2)$ approaching $(0,0)$, and for all choices the energy spectrum is determined by (\ref{Kiomhalf}).

The boundary condition (\ref{MIT1}) implies the vanishing normal probability current and scalar densities at the boundary $\xi=0$, namely,
\begin{eqnarray}
  \left. i  J^{\rm D}_N \right|_{\xi=0}  
	\equiv \left. i N_\mu \bar{\psi}^{\rm D}
	\gamma^\mu_{\rm R} 
	\psi^{\rm D}
	\right|_{\xi=0} 
	&=& \left.\bar{\psi}^{\rm D}
	\psi^{\rm D}
	\right|_{\xi=0} \nonumber\\
  &=& \left.-\bar{\psi}^{\rm D}
	\psi^{\rm D}
	\right|_{\xi=0} \nonumber\\
	&=& 0~.	\label{BCA}
\end{eqnarray}
The first line is (\ref{MIT1}) multiplied by $\bar{\psi}$ on the left, the second line is the Hermitian conjugate of (\ref{MIT1}) multiplied by $\gamma^0 \psi$ on the right, and the third line results from the equality of the first and the second lines. Examples of the normal current density $J^{\rm D}_N$ and scalar density $\bar{\psi}\psi$ of the Dirac bouncers at the lowest few eigen-energies will be shown in Sec. \ref{compareBP}. 

At the scale of our interest, there is no need of considering the boundary condition in the chiral (or little/cloudy) bag model \cite{Rho1,Rho2,CT75,Rho3,Theberge, Theberge2, Thomas, Hosaka, Tsushima}, which is more realistic in nuclear physics. Indeed, the original bag models describe how quarks confined in a hadron interact, and the scale of the thickness of the bag boundary is supposed to be much less than 1 fm. But here we are looking at the surface of a mirror, or the boundary of a medium for the UCN. So the boundary thickness here is of the atomic scale $\sim 0.1$ nm, which is much greater than the size of the whole hadron (the bag), not to mention the boundary of the hadron.

More specifically, the chiral bag model is proposed to handle the residual strong interaction outside the bags and between the nucleons, with the approximated chiral symmetry of light quarks in the bags taken into account. 
At the scale of nucleon-nucleon interactions, which is larger than the scale of a bag, 
the approximated chiral symmetry in a high-energy Lagrangian 
has been explicitly broken in the low-energy effective field theory.
As in the chiral perturbation theory (ChPT) \cite{Epelbaum}, the physics in the truncated higher-orders terms 
relevant to the approximated chiral symmetry are not resolvable by the hadron fields in the effective theories at the lowest orders. Those higher-order physics are encoded in the parameters of the effective theories such as the coupling constants and scattering lengths. The values of those parameters may be input from experiments, while they are extremely difficult to be obtained
from first-principle (QCD) calculations.

The leading order from ChPT is the one-pion-exchange Yukawa potential supplemented by contact interactions (delta-function potentials) \cite{Weinberg90, ORvK96, EM13}. 
For cold neutrons with de Broglie wavelengths of nm scale, even the Yukawa potential of a nucleon (significant within a few fm from the nucleus surface) is not resolvable. A single atomic nucleus would be experienced like a point particle by these cold neutrons and so described as a contact potential of delta function multiplied by the scattering length in the effective theory at this scale \cite{Squires78}.
The Fermi pseudo-potential of the medium consisting of the atoms is then approximated as the sum of all the delta-function contact potentials of the atomic nuclei in the medium. From the Fermi pseudo-potential one can derive a macroscopic refractive index for the medium. From this wavelength-dependent refractive index, one can see that cold neutrons can be totally reflected by a medium at any incident angle if their wavelengths are sufficiently long \cite{Squires78, IMFrank}.

The wavelengths of such UCN are no less than about 50 nm, which are greater than the lattice constant of a solid medium or the averaged intermolecular distance in a liquid. So one can further coarse-grain the Fermi pseudo-potential and write down an effective potential as a finite barrier to describe this macroscopic reflection. Then one ends up with the conventional effective quantum-mechanical theory for the UCN reflection. Our model is a further idealization of this kind of the conventional effective theory: The conventional finite barrier of the Fermi pseudo-potential is replaced by the boundary condition of the BC-MIT in order to obtain analytic results while avoiding the Klein paradox caused by an infinite potential barrier. We apply the mathematical form of this boundary condition for our purpose, while the physics described here are at a scale very remote from those of the original MIT and chiral bag models.

\subsection{Majorana particles in Rindler coordinates}

Given the positive-energy modes $\tilde{\psi}_{\omega{\bm {\bm k_\perp}} \sigma}^{\rm D+}$ of Dirac particles, 
a wave packet of Majorana particle in the Majorana representation in Rindler coordinates can be constructed as a real solution for Dirac equation 
\begin{equation}
\tilde{\psi}^{\rm M}(\eta,\xi,{\bm x_\perp})=
  \sum_{\sigma=\pm} \int_{-\infty}^{\infty}\frac{d{\bm {\bm k_\perp}}}{2\pi} \int_0 ^{\infty} d\omega 
	\left[ b(\omega, {\bm {\bm k_\perp}},\sigma)
  \tilde{\psi}_{\omega{\bm {\bm k_\perp}} \sigma}^{\rm D+}
	(\eta,\xi,{\bm x_\perp}) + 
   b^*(\omega, {\bm {\bm k_\perp}},\sigma) \left(\tilde{\psi}_{\omega{\bm {\bm k_\perp}} \sigma}^{\rm D+}
	 \right)^C (\eta,\xi,{\bm x_\perp})\right],
\label{Majorana}
\end{equation}
where $b(\omega, {\bm {\bm k_\perp}},\sigma)$ is the amplitude, and $\tilde{\psi}^C = i \tilde{\gamma}^2 \tilde{\psi}^*$ is the charge conjugate of $\tilde{\psi}$. 
In particular, a single-mode Majorana wave function in the Dirac representation reads \cite{Gardner16}\footnote{The most general 
condition that a Majorana wave function should satisfy is $(\psi^{\rm M})^C=\pm \psi^{\rm M}$ \cite{Gardner16}, but here we only adopt the condition (\ref{MajoranafromDirac}) where the charge conjugation of a Majorana wave function is exactly the same as itself.}
\begin{eqnarray}
\psi_{\omega {\bm {\bm k_\perp}}\sigma}^{\rm M} 
&=&\frac{1}{\sqrt{2}}\left[ \psi_{\omega {\bm {\bm k_\perp}}\sigma}^{\rm D+} + \left(\psi_{\omega {\bm {\bm k_\perp}}\sigma}^{\rm D+}\right)^C \right]
=\left(\psi_{\omega{\bm {\bm k_\perp}} \sigma}^{\rm M}\right)^C  
\label{MajoranafromDirac}
\end{eqnarray}
which satisfies the same boundary condition as $\psi_{\omega {\bm {\bm k_\perp}}\sigma}^{\rm D+}$ does. 

Inserting the Dirac wave function $\psi_{\omega {\bm {\bm k_\perp}}\sigma}^{\rm D+}$ in (\ref{Diracspinor4dim}) into (\ref{MajoranafromDirac}), we get the Majorana wave function $\psi_{\omega {\bm {\bm k_\perp}}\sigma}^{\rm M}$.
Then, imposing the BC-MIT (\ref{MIT1}) into Majorana wave function (\ref{MajoranafromDirac}), 
we obtain the condition 
\begin{eqnarray}
&&F(\eta, {\bm x_\perp}) \left\{ \left[ \kappa K_+(0)+ \left(m-ik_1-k_2\right)K_-(0) \right] A_+ 
+ \left[i\kappa K_-(0) +i\left( m -i k_1- k_2\right)K_+(0)\right]A_-\right\}\nonumber\\
&&+ F^*(\eta, {\bm x_\perp}) \left\{ \left[i\kappa K_-(0)+i\left(m +i k_1+ k_2\right) K_+(0)\right] A_+^* +
\left[-\kappa K_+(0) -\left(m + ik_1 +k_2\right) K_-(0)\right] A_-^*\right\}=0,
\label{boundstatesMajorana4} 
\end{eqnarray}
where $F(\eta, {\bm x_\perp}) \equiv e^{-i\omega\eta}e^{i{\bm k_\perp}\cdot{\bm x_\perp}}/(im+k_1)$
(note that $K_-(0) = K_{\frac{i\omega}{a}-\frac{1}{2}}(\kappa/a)= K^*_{\frac{i\omega}{a}+ \frac{1}{2}}(\kappa/a) = K_+^*(0)$).
Since (\ref{boundstatesMajorana4}) has to be satisfied for all $\eta$ and ${\bm x_\perp}$, one must have the $F$ terms and the $F^*$ terms vanishing separately, which lead to Eq. (\ref{boundstatesDirac4}) again. 
Thus, the eigen-energies of a Majorana bouncer are the same as those of a Dirac bouncer. 
The energy levels of both Dirac and Majorana 
bouncers are independent of the spin orientation determined by $A_\pm$. 
Equation~(\ref{BCA}) derived from (\ref{MIT1}) is still true for Majorana particles. 
However, since a Majorana particle is its own antiparticle,
their scalar density $\bar{\psi}\psi$ vanishes everywhere, while the normal probability current density $J_N$  does not.

In Sec.~\ref{compareBP},  we restrict our attention to the special case of ${\bm k_\perp}=0$ (normal incidence) for simplicity.

\section{Comparison of normally incident bouncing particles of different nature}
\label{compareBP}

\subsection{Energy levels and transition frequencies}
Let us compare the energy levels of the above four kinds of the bouncing particles of ${\bm k_\perp}=0$.

\paragraph{Nonrelativistic particles.}
The energy levels ${\cal E}^{\rm NR}_n(\equiv E_n)$ obtained from the Schro\"{o}dinger equation with the ideal boundary condition (\ref{energylevel}) are 
\begin{equation}
\mathcal{E}_n^{\rm NR}=\left(\frac{m a^2}{2}\right)^{1/3}\zeta_n, 
\hspace{1cm} n=1,2,3,\ldots \label{EnNR}
\end{equation}
where $a$ is the uniform (gravitational) acceleration and $-\zeta_n$ is the $n$-th zero 
of the Airy function given by ${\rm Ai}(-\zeta_n)=0$.

\paragraph{KG particles.}
The energy levels $\mathcal{E}_n^{{\rm KG}}$ from the KG equation in Rindler coordinates are given by the condition
(\ref{Kiom}), which implicitly assumes an infinite potential barrier as the ideal floor/mirror. 
The formula for the approximated value of $\omega_n/a$ when the fixed $\mu$ in the condition  
$K^{}_{i\omega_n/a}(\mu)=0$ is right at the $n$th zero of the modified Bessel function has been given in Ref.~\cite{Ferreira} as
\begin{equation}
  \frac{\omega_n}{a} \approx \mu + \zeta_n 2^{-1/3} \mu^{1/3} + \frac{\zeta_n^2}{60}\,2^{1/3} \mu^{-1/3}
	+ \left( \frac{1}{70} - \frac{\zeta_n^3}{700} \right) \mu^{-1} + O\left(\mu^{-5/3}\right)
\label{zerosmodifiedbessel}
\end{equation}
for large $\omega_n/a$ and small $\zeta_n$, with the $n$th zero of the Airy function $-\zeta_n$.
To compare with $\mathcal{E}_n^{\rm NR}$ in (\ref{EnNR}),
one should exclude the rest-mass energy $mc^2$ $(c\equiv 1)$ from $\hbar \omega$ $(\hbar\equiv 1)$, namely, 
define the subtracted energy ${\cal E}^{\rm KG}\equiv \omega-m$ such that the relativistic correction to the eigen-energy is 
\begin{equation}
	\Delta {\cal E}_n^{\rm KG} = {\cal E}_{n}^{\rm KG} - \mathcal{E}_n^{\rm NR}, 
\label{energyRindlerspace}
\end{equation}
where $n$ labels the $n$th energy level. From Eq. (\ref{zerosmodifiedbessel}) 
with $\mu = \kappa/a=m/a$ and $\nu_n = i\omega_n/a$ ($a=g$) 
we find an explicit expression of the first-order correction to the energy as
\begin{eqnarray}
 \Delta{\cal E}_n^{\rm KG} \approx {\cal E}_n^{(1)} \equiv 
   a \frac{\zeta_n^2}{60}\,2^{1/3} \mu^{-1/3} = 
   m a \mathcal{B} \left(\frac{2a}{m}\right)^{2/3} \frac{\zeta_n^2}{60}.
	\label{KGEnCorrection}
\end{eqnarray}
The above correction is positive and goes to zero in the large-mass and small-acceleration limit, $a/m \rightarrow 0$. 
It can be recovered in a perturbative analysis of the Hamiltonian in Appendix~\ref{perturbation}.

In the example shown in Table \ref{tableeigenvalues}, one can see that the $n$th energy level of a KG bouncer 
is higher than the energy level of a nonrelativistic bouncer 
with the same $n$ and all other parameters, and the correction is growing as $n$ increases. 
This is consistent with the behavior of our approximated result in (\ref{KGEnCorrection}).

\paragraph{Dirac and Majorana particles.}

The energy levels $\mathcal{E}_n^{\rm D,M}$ for Dirac and Majorana bouncers,
 respectively, with ${\bm {\bm k_\perp}}=0$ from the Dirac equation in Rindler coordinates are determined by (\ref{Kiomhalf}) with the solutions $\omega=\omega_n = \mathcal{E}_n^{\rm D,M}+m$, $n=1,2,3,\cdots$. Following the same method in Ref.~\cite{Ferreira}, the values of $\omega_n/a$ can be expressed as an asymptotic expansion in $\mu\equiv m/a$, 
\begin{equation}
  \frac{\omega_n}{a} \approx \mu - \frac{1}{2} + \zeta_n 2^{-1/3}  \mu^{1/3} + \frac{\zeta_n^2}{60}\,2^{1/3} \mu^{-1/3}
	+ \frac{\zeta_n}{6}\,2^{-1/3}\mu^{-2/3} + \left( \frac{1}{70} - \frac{\zeta_n^3}{700} -\frac{1}{12} \right) \mu^{-1} + 
	O\left(\mu^{-4/3}\right)
	\label{zerosBCMIT}
\end{equation}  
for large $\omega_n/a$ and small $\zeta_n$.
Comparing (\ref{zerosBCMIT}) with the expansion (\ref{zerosmodifiedbessel}) from the Dirichlet boundary condition for KG  bouncers,
one can see that the BC-MIT gives three extra terms to Dirac bouncers:
$-1/2$, $-\mu^{-1}/12$, and $(\zeta_n/6)\,2^{-1/3}\mu^{-2/3}$, up to the order of $\mu^{-1}$.
The first two terms are independent of $\zeta_n$ and negative. They equally shift down all the energy levels, and make the lowest few energy levels of Dirac bouncers lower than the ones of nonrelativistic bouncers and KG bouncers (e.g. energy levels of $n\le 3$ in Table \ref{tableeigenvalues}, where we chose a modest value of $m/a = 10$ to demonstrate the effects of the corrections). This negative energy shift will be overcome by the positive $\zeta_n$-dependent corrections in (\ref{zerosBCMIT}) when $n$ gets sufficiently large. 

Nevertheless, a uniform shift in energy spectrum is not detectable in laboratories. One detectable quantity is the transition frequency between (the $n$th and the $n'$th) energy eigenstates (e.g., the Gravity Resonance Spectroscopy (GRS) in Refs.~\cite{Abele, Jenke, Sedmik, Cronenberg}),
\begin{eqnarray}
  \omega_{n',n}=\mathcal{E}^{}_{n'} -\mathcal{E}^{}_{n}.
\end{eqnarray}
From (\ref{zerosmodifiedbessel}) and (\ref{zerosBCMIT}), the relativistic correction to the transition frequency between the $n$th and the $n+1$th eigenstates for a KG bouncer is approximately
\begin{equation}
  \Delta\omega_{n+1,n}^{\text{KG}} = \omega_{n+1,n}^{\text{KG}} -\omega_{n+1,n}^{\text{NR}} \approx 
	\frac{1}{60}\left(\frac{2 a^4}{m}\right)^{1/3}\left(\zeta_{n+1}^2-\zeta_n^2\right), \label{DomegaKG}
\end{equation}
which is always positive and grows as $n$ increases while $a$ and $m$ are fixed. Similar correction for a Dirac or Majorana bouncer is given as
\begin{equation}
  \Delta\omega_{n+1,n}^{\text{D,M}} \approx \frac{1}{60}\left(\frac{2 a^4}{m}\right)^{1/3}\left(\zeta_{n+1}^2-\zeta_n^2\right) 
	+\frac{1}{6} \left(\frac{a^5}{2 m^2}\right)^{1/3}\left(\zeta_{n+1}-\zeta_n\right), \label{DomegaDM}
\end{equation}
which is also positive.
In Table \ref{tabletransitionProba1} we list the transition frequencies of the neighboring energy levels from Table \ref{tableeigenvalues}. 
One can see the above tendency at every $n$. A comparison of the lowest few (scaled) energy levels for the KG and Dirac bouncers can also be found in Figs.~1 and 2 of Ref.~\cite{Boulanger}.

Since neutron mass is $m \approx 0.94$ GeV \cite{PD20}, and the gravitational acceleration on the Earth surface is $a = 9.8\,{\rm m/s}^2 \approx 2.15 \times 10^{-32}$ GeV, one has a very large value of the parameter $\mu=m/a \sim 4.37 \times 10^{31}$, implying that (\ref{zerosmodifiedbessel}) and (\ref{zerosBCMIT}) are good approximations for realistic bouncing UCN experiments. The ratio of the $-1/2$ correction in (\ref{zerosBCMIT}) for Dirac and Majorana bouncers to $\mathcal{E}_n^{\rm NR}/a$ is about $10^{-11}$, which is very small here. Inserting these parameter values into (\ref{DomegaKG}) and (\ref{DomegaDM}), one finds that the relativistic corrections to transition frequencies, $\Delta\nu_{n+1,n}^{\text{KG}} = \Delta\omega_{n+1,n}^{\text{KG}}/(2\pi)$ and $\Delta\nu_{n+1,n}^{\text{D,M}}$, both are of the order of $10^{-20}$ Hz  (cf. $\nu^{\text{NR}}_{n+1,n} \approx 254.45$, $208.32$, $184.11$, $168.30$, $156.83$ Hz for $n=1,2,3,4,5$) for the lowest energy levels and way too small to be detected by current technology.

In the proposals of similar quantum bouncer experiments but alternatively using Rydberg Ps atoms instead of the UCN, the internal energy of each Ps atom is designed to be highly excited to extend the lifetime of the Ps atom. 
For example, the authors of Ref. \cite{Crivelli2015} argued that their Ps atom with the principal quantum number $n'$ for the electron-positron interaction should be prepared at $n' > 30$ to make the lifetime of the atom longer than the observation time. Nevertheless, the total energy of a slow Ps atom is dominated by its rest mass, which is about two times of electron mass ($2 m_e c^2 \approx 1$ MeV) and roughly $10^{-3}$ times of neutron mass. Indeed, the energy difference between a Rydberg Ps atom and a Ps atom in the ground state is no greater than their ionization energy, which is of the order of $10^{-3}$ times of $13.6$ eV for a hydrogen atom, much smaller than 1 MeV. Recall that the leading orders of the energy-level differences from (\ref{zerosmodifiedbessel}) and the relativistic corrections to transition frequency in (\ref{DomegaKG}) for a KG bouncer are proportional to $m^{1/3}$ and $m^{-1/3}$, respectively. Substituting $m=2m_e$ to (\ref{zerosmodifiedbessel}) 
\footnote{It may be doubtful to apply the KG equation to ortho-positronium atoms, which are composite bosons of total spin $S=1$ rather than spinless. Extrapolating from (\ref{zerosmodifiedbessel}) and (\ref{zerosBCMIT}), however, we expect that the dimension of our estimate would be correct, and the values of our estimate would be good up to order of magnitude.}, one can see that the energy levels of a Ps atom in a gravitational field are about 10 times denser than the UCN's, and so the transition frequencies for the lowest energy levels are of the order of 10 Hz. As for the relativistic corrections to the transition frequencies of Ps atoms, (\ref{DomegaKG}) with $m=2m_e$ implies that the corrections are about 10 times of the UCN's in value, namely, of the order of $10^{-19}$ Hz for the lowest energy levels. While the contrast between the relativistic correction and transition frequency will be improved by a factor of $10^2$ if we replace the UCN by Ps atoms, the relativistic corrections for quantum Ps bouncers will still be way below the resolution of current technology.

\begin{table}
  \begin{center}
\caption{Comparison of the lowest six energy levels of the nonrelativistic (left column), KG (middle), Dirac, and Majorana bouncers 
 (right) for $\mu=m/a=10$. Here each $\mathcal{E}_n$ is scaled by a factor $ma\mathcal{B}(=(ma^2/2)^{1/3})$.}
\begin{tabular}{ccccc}
  \hline
  \hline
  $~~n~~$&~~$\mathcal{E}_n^{\text{NR}}/ma\mathcal{B}$~~&~~$\mathcal{E}_n^{\text{KG}}/ma\mathcal{B}$~~&~~$\mathcal{E}_n^{\text{D,M}}
  /ma\mathcal{B}$  \\
  \hline
   1& 2.338 & 2.369 & 2.103 
\\ 2& 4.088 & 4.179 & 3.931 
\\ 3& 5.521 & 5.683 & 5.446
\\ 4& 6.787 & 7.028 & 6.800
\\ 5& 7.944 & 8.270 & 8.049
\\ 6& 9.023 & 9.438 & 9.223
\\
\hline
\end{tabular}
\label{tableeigenvalues}
\end{center}
\end{table}
\begin{table}
  \begin{center}
\caption{
Scaled transition frequencies between the $n$th and ($n+1$)th energy eigenstates of the bouncersin Table \ref{tableeigenvalues} for $m/a=10$.}
\begin{tabular}{cccccc}
  \hline
  \hline
  $~~n~~$&~~$\omega_{n+1,n}^{\text{NR}}/ma\mathcal{B}$~~&~~$\omega_{n+1,n}^{\text{KG}}/ma\mathcal{B}$~~&
	~~$\omega_{n+1,n}^{\text{D,M}}/ma\mathcal{B}$ \\ 
  \hline
   1& 1.750 & 1.810 & 1.828 
\\ 2& 1.433 & 1.504 & 1.515 
\\ 3& 1.266 & 1.345 & 1.354 
\\ 4& 1.157 & 1.242 & 1.249 
\\ 5& 1.079 & 1.168 & 1.174 
\\
\hline
\end{tabular}
\label{tabletransitionProba1}
\end{center}
\end{table}

\subsection{Probability density}

\begin{figure}[t!]
 \begin{center}
\includegraphics[width=15cm]{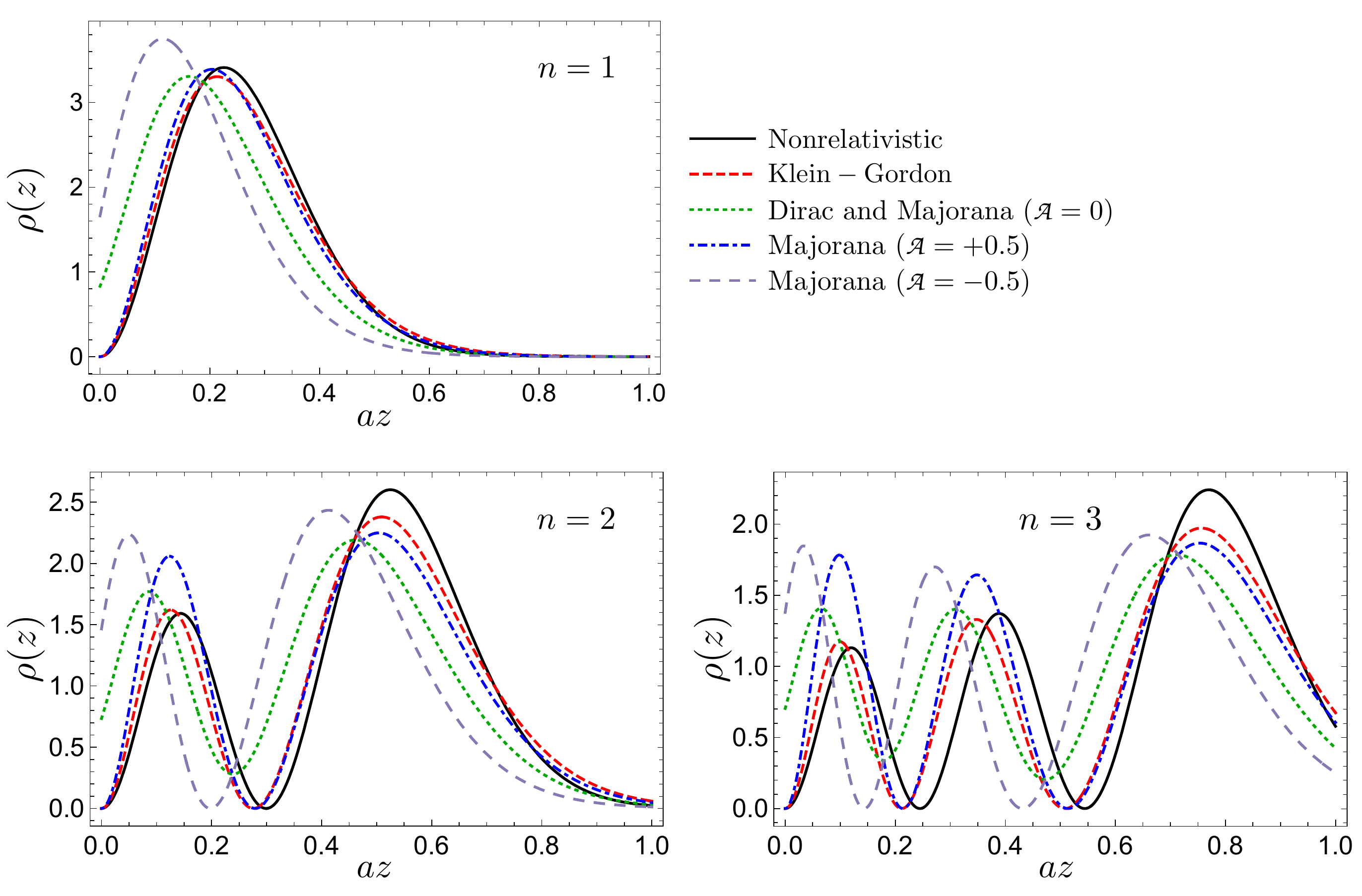}
\caption{\label{probabilitydensity} 
The normalized probability densities $\rho(z)$ of the ground, first-excited and second-excited states, with the transverse dimensions $x$ and $y$ suppressed. Here, $m/a=10$, $az=e^{a\xi}-1$, 
and we show three cases of various values of $\mathpzc{A}$ for Majorana bouncers. $\mathpzc{A}=0$ is for the cases of $A_+=\pm iA_-$, corresponding to the spin orientations in the $\pm z$-directions. 
$\mathpzc{A}=\pm 0.5$ are the upper and lower limits of $\mathpzc{A}=+0.5\sin (2\omega\eta)$ and $\mathpzc{A}=-0.5\sin (2\omega\eta)$ from purely real coefficients $A_\pm$ with $A_+=A_-$ and purely imaginary $A_\pm$ with $A_+=A_-$, respectively.
}
\end{center}
\end{figure} 

Let us compare the probability density $\rho(z)$ with the transverse dimensions $x$ and $y$ suppressed 
for the bound states of different bouncing particles of ${\bm k_\perp}=0$.

\paragraph{Nonrelativistic particles.}
The probability density for a nonrelativistic bouncer is given by normalizing
\begin{eqnarray}
\rho^{{\rm NR}}(z)&=&\psi^*(z)\psi(z)\nonumber\\
&= & \mathcal{N}^2_n\left[{\rm Ai}\left(\frac{z}{\mathcal{B}}-\zeta_n\right)\right]^2,
\end{eqnarray}
from (\ref{NRwavefunc}). Here, $\mathcal{B}$, $\mathcal{N}_n$, and $\zeta_n$ are 
given in Eqs.~(\ref{beta}), (\ref{normalizationfactor}), and (\ref{energylevel}), respectively.

\paragraph{KG particles.}
The probability density for a KG bouncer is given by normalizing
\begin{eqnarray}
\rho^{{\rm KG}}(z)&=&i\bigg[\phi^*(z)\frac{\partial \phi(z)}{\partial \eta}-\phi(z)\frac{\partial \phi^*(z)}{\partial \eta}\bigg]\nonumber\\
&=& \frac{\omega\sinh(\pi\omega/a)}{2\pi^4a}\left[ K_{\frac{i\omega}{a}}(Z)
\right]^2,
\end{eqnarray} 
from (\ref{KGmodeFree}) with $\omega_n=\mathcal{E}_n^{{\rm KG}}+m$ determined by the boundary condition (\ref{Kiom}), 
and $Z \equiv me^{a\xi}/a \equiv m(az+1)/a$. Here, $K_{\frac{i\omega}{a}}(Z)$ is a real function of $Z$ and we adopt the normalization condition $\int^\infty_0 \rho^{\rm KG}(z) e^{a\xi}d\xi=1$.

\paragraph{Dirac particles.}
The probability density for a positive-energy Dirac bouncer of ${\bm {\bm k_\perp}}=0$ is given by normalizing
\begin{eqnarray}
\rho^{\rm D}_{\omega\sigma}(z)&=&n_0\bar\psi^{\rm D}_{\omega\sigma}(z)\gamma_{\rm R}^0\psi^{\rm D}_{\omega\sigma}(z)\nonumber\\
&=& 8\left|{\cal N}_{\omega\sigma}^{\rm D}\right|^2 (|A_+|^2+|A_-|^2) 
 \left| K_{\frac{i\omega}{a}-\frac{1}{2}}(Z) \right|^2, 
\label{probDirac}
\end{eqnarray}
from (\ref{Diracspinor4dim}) with $\omega=\mathcal{E}_n^{{\rm D,M}}+m$ determined by the boundary condition (\ref{Kiomhalf}), where $n_\mu$ is the normal vector perpendicular to the constant time hypersurface. We have used $K_{\frac{i\omega}{a}+\frac{1}{2}}(Z)= K^*_{\frac{i\omega}{a}-\frac{1}{2}}(Z)$ for real $Z$, and noticed that $\partial_\mu ( \sqrt{-g}~\bar{\psi}\gamma_{\rm R}^\mu \psi) = \sqrt{-g}~\nabla_\mu ( \bar{\psi}\gamma_{\rm R}^\mu \psi ) = 0$. Here, the normalization constant ${\cal N}^{\rm D}_{\omega\sigma}$ is determined by the condition $\int^\infty_0 \rho^{\rm D}_{\omega\sigma}(z) e^{a\xi} d\xi=1$.

Our result of $\rho^{\rm D}_{\omega\sigma}$ in (\ref{probDirac}) is independent of time $\eta$. 
The overall factor $(|A_+|^2+|A_-|^2)$ in (\ref{probDirac}) is real and can be normalized while the coefficients $A_\pm$ determining the 
Dirac bouncer's spin orientation are complex numbers. So the above probability density of a 
Dirac bouncer with ${\bm {\bm k_\perp}}=0$ is independent of its spin orientation.

For a single-mode, positive-energy Dirac bouncing particle with general ${\bm {\bm k_\perp}}$, one still has $\rho^{\rm D}_{\omega\sigma}(z) \propto | K_{\frac{i\omega}{a}-\frac{1}{2}}(Z) |^2$, and only the overall factor depends on $A_\pm$ and ${\bm k}$, which can be normalized, too. We conclude that the probability density of a single-mode Dirac bouncer of positive-energy is static and independent of its spin orientation in general. 

\paragraph{Majorana particles.} 
The probability density for a Majorana bouncer of ${\bm {\bm k_\perp}}=0$ is given as
\begin{eqnarray}
\rho^{{\rm M}}_{\omega\sigma}(z) &=&n_0\bar\psi^{\rm M}_{\omega\sigma}(z)\gamma_{\rm R}^0\psi^{\rm M}_{\omega\sigma}(z) \nonumber\\
&=& 8\left|{\cal N}_{\omega\sigma}^{\rm M}\right|^2\left(|A_+|^2+|A_-|^2\right) 
    \left\{ \left|K^{}_{\frac{i\omega}{a}-\frac{1}{2}}(Z)\right|^2 
    +2 \mathpzc{A}(\eta)\,{\rm Re}\left[\left(K_{\frac{i\omega}{a}-\frac{1}{2}}(Z)\right)^2\right]
		\right\},
\label{probMajorana}
\end{eqnarray}
from (\ref{MajoranafromDirac}) with 
$\omega=\mathcal{E}_n^{{\rm D,M}}+m$ determined by the boundary condition (\ref{Kiomhalf})  
shared with Dirac particles. Here, ${\cal N}_{\omega\sigma}^{\rm M}$ is the normalization constant of Majorana wave function in the case of ${\bm k}_\perp=0$, which is determined by the condition $\int^\infty_0 \rho^{\rm M}_{\omega\sigma}(z) e^{a\xi}d\xi=1$. In contrast to Dirac bouncers, the behavior of the probability density for a Majorana bouncer
depends on spin orientation (determined by the values of $A_\pm$) in the factor 
\begin{eqnarray}
  \mathpzc{A}(\eta) \equiv
	\frac{-{\rm Im}\left[\left(A_+^2+A_-^2\right)e^{-2i\omega\eta}\right]}
  {2\left(|A_+|^2+|A_-|^2\right)}. \label{bigA}
\end{eqnarray}
When $A_+= \pm iA_-$ (corresponding to spin orientation in $\pm z$-directions \cite{Crispino, Matsas, Vanzella}), one has $\mathpzc{A}=0$ and the probability density for the Majorana bouncer of ${\bm {\bm k_\perp}}=0$ is identical to the probability density for its Dirac counterpart. In other spin orientations, $A_+^2+A_-^2 \not= 0$, and so $\mathpzc{A}(\eta)$ oscillates at a very high frequency $2\omega$ ($> 2 m \sim 10^{23}$ Hz for neutrons). This corresponds to {\it Zitterbewegung} from the interference between the positive- and negative-energy components of the Majorana wave function (\ref{MajoranafromDirac}) \cite{Greiner00}. In the cases with significant {\it Zitterbewegung}, the one-particle interpretation of the wave function does not work, and the quantity $\rho^{{\rm M}}_{\omega\sigma} \sim \psi^{\rm M\,\dagger}_{\omega \sigma}\psi^{\rm M}_{\omega\sigma}$ should not be interpreted as a probability density. Anyway, at the scale of our interest, such a spin-dependent fast oscillation would not be resolvable by an apparatus, which may find the factor $\mathpzc{A}$ vanishing since the time-average of it is zero. Then the probability density distribution of a Majorana bouncer appears no difference from its Dirac counterpart.

In FIG. \ref{probabilitydensity}, we show the probability densities for KG, Dirac, and Majorana bouncers around the boundary at $\xi=0$ 
to compare with the nonrelativistic case of the same parameter values. 
Similar to the nonrelativistic case, the wave functions for KG particles vanish at the ideal mirror boundary \cite{Saa,Boulanger}, 
and so the probability of finding the KG particle right on the floor is zero. In contrast, 
the probability density for a Dirac particle does not vanish on the boundary, associated with the lowest energy levels of the Dirac bouncer lower than their nonrelativistic limits,
while its scalar density vanishes on the boundary. The behavior of the time-averaged probability density for a Majorana bouncer is the same as its Dirac counterpart.

\subsection{Scalar density}

\begin{figure}[t!]
  \begin{center}
\includegraphics[width=8cm]{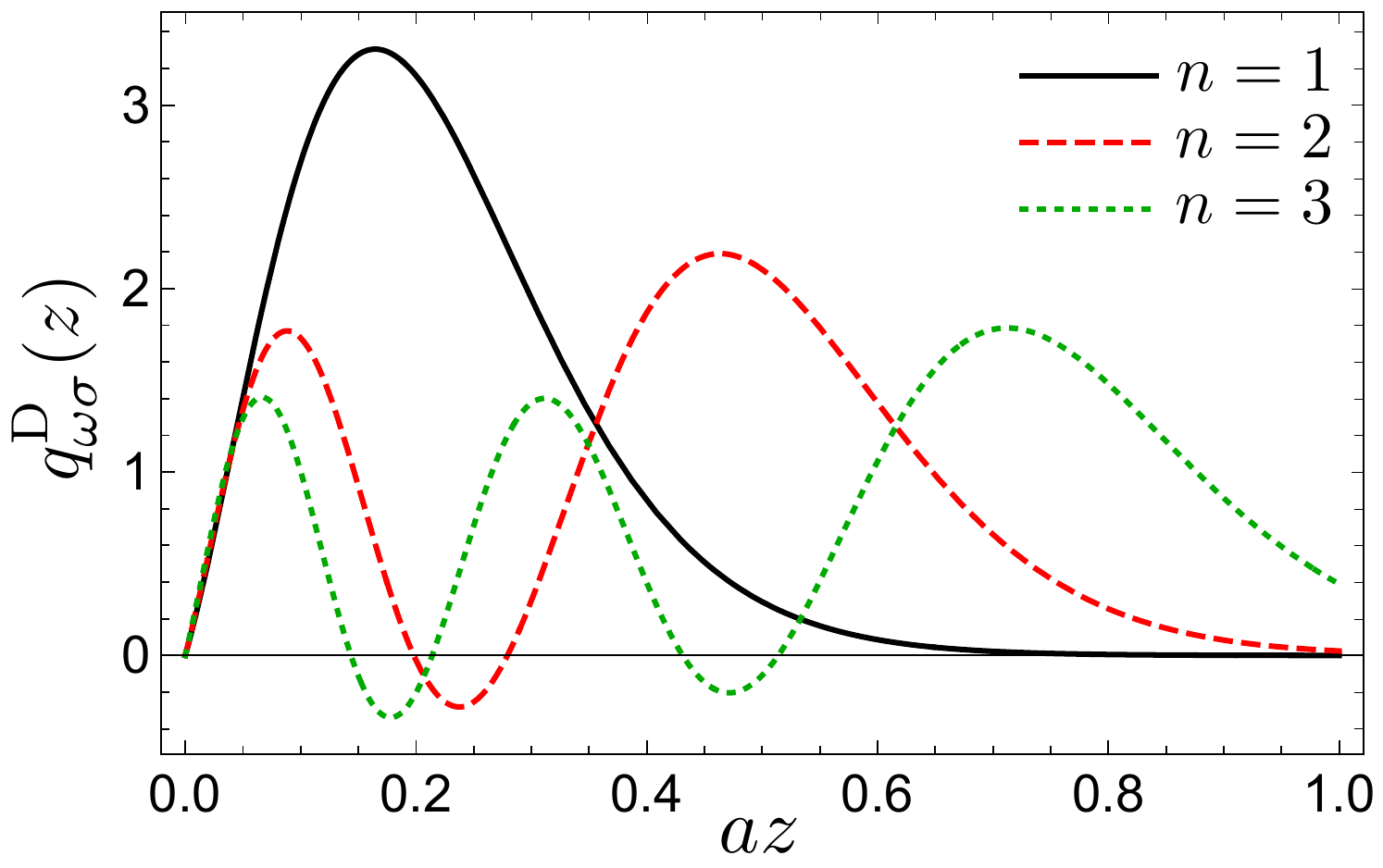}
\caption{\label{chargedensity} 
The scalar density $q^{\rm D}_{\omega\sigma}(z)$ of a Dirac bouncer of ${\bm k_\perp}=0$ in the ground, first-excited, and second-excited states. 
Here, $m/a=10$ and $az=e^{a\xi}-1$. } 
\end{center}
\end{figure} 

The scalar density for the Majorana wave function (\ref{Majorana}) 
is exactly zero everywhere since its amplitude $b$ of the positive-energy part is exactly the complex conjugate of the amplitude of the negative-energy part.
The scalar density for the positive-energy wave function (\ref{Diracspinor4dim}) of Dirac bouncer in the case of ${\bm k_\perp}=0$ reads
\begin{equation}
q^{\rm D}_{\omega\sigma}(\xi) = \bar{\psi}_{\omega\sigma}^{\rm D}\psi_{\omega\sigma}^{\rm D} 
=8\left|{\cal N}_{\omega\sigma}^{\rm D}\right|^2 
\left(|A_+|^2+|A_-|^2\right) {\rm Im}\left\{\left[K_{\frac{i\omega}{a}+\frac{1}{2}}(Z)\right]^2 \right\}.
\label{chargedensityDirac2Dim}
\end{equation}
In FIG.~\ref{chargedensity}, we show the above scalar density for the ground, first-excited, and second-excited states of a Dirac bouncer. 
All vanish at $z=0$ ($\xi=0$), as required by the BC-MIT.

\subsection{Normal probability current density}

The normal probability current density 
$J^{\rm D}_{N, \omega\sigma}= \bar\psi^{\rm D}_{\omega\sigma}\gamma^3\psi^{\rm D}_{\omega\sigma}$
for the wave function (\ref{Diracspinor4dim}) of a Dirac bouncer of ${\bm {\bm k_\perp}}=0$ is exactly zero everywhere, 
while the normal probability current density for its Majorana counterpart in the state (\ref{MajoranafromDirac}) with ${\bm k_\perp}=0$
\begin{eqnarray}
J^{\rm M}_{N, \omega\sigma}(z) &=& \bar\psi^{\rm M}_{\omega\sigma}\gamma^3
  \psi^{\rm M}_{\omega\sigma} \nonumber\\
&=& 8\left|{\cal N}_{\omega\sigma}^{\rm M}\right|^2 
{\rm Re}\bigg[ \left(A_-^2 + A_+^2\right) e^{-2i \omega \eta}\bigg]
{\rm Im}\left[\left(K_{\frac{i\omega}{a}+\frac{1}{2}}(Z)\right)^2\right]
\label{CurrentMajorana2Dim}
\end{eqnarray}
is zero at $z=0$ by the BC-MIT but in general nonvanishing for $z>0$ and oscillating in time, except the cases of $A_+= \pm iA_-$. This spin-dependent {\it Zitterbewegung}
would be coarse-grained by an apparatus of slow response, which would detect a vanishing time-averaged $J^{\rm M}_{N, \omega\sigma}$.

Our result does not imply that the reflection probability of a Majorana bouncer depends on its spin orientation. 
The above normal probability current density $J^{\rm M}_{N, \omega\sigma}$ 
is vanishing at the boundary $z=0$ for all time, thus the Majorana particle is perfectly reflected, and its reflection probability is $100\%$ for all spin orientations.
Note that, unlike those in the scattering problems of free particles from/to the asymptotic regions, each of our bound-state wave functions (\ref{Diracspinor4dim}) and (\ref{MajoranafromDirac}) in uniformly accelerated frames is associated with a unique spin orientation, and so the reflected particles in these states have the same spin orientation as the incident ones.

The dependence of spin orientation in {\it Zitterbewegung} has nothing to do with parity violation. Although neither the Dirac particles nor Majorana particles considered in this paper are parity eigenstates, the BC-MIT we introduced does respect parity symmetry both in the systems of Dirac and Majorana particles. This is clear in the Lagrangian of the MIT bag model (e.g., Refs.~\cite{Hosaka} and \cite{CT75}), which is invariant under parity transformations $\psi^{\rm D}(t, {\bm x})$ to $\hat{P} \psi^{\rm D}(t, {\bm x}) \hat{P}^\dagger = \gamma^0 \psi^{\rm D}(t, -{\bm x})$ and $\psi^{\rm M}(t, {\bm x})$ to $\hat{P} \psi^{\rm M}(t, {\bm x}) \hat{P}^\dagger = i \gamma^0 \psi^{\rm M}(t, -{\bm x})$,
where $\hat{P}$ is the unitary operator of parity transformation.

\section{Summary and Conclusion}
\label{SummaryConclusion}

We have studied the relativistic effect on the quantum states of a particle trapped in a gravitational potential and bouncing above a floor. 
According to the equivalence principle of relativity, we revisited the KG and Dirac equations in Rindler coordinates under appropriate boundary conditions. As has been shown in Ref.~\cite{Boulanger}, the energy levels reduce to the well-known formula in the nonrelativistic limit, which is given by the Schr\"{o}dinger equation with a gravitational potential linear in altitude above a floor of ideal mirror. Our results verify that the relativistic correction from the KG equation with the Dirichlet boundary condition of an ideal-mirror floor raises each energy levels from the nonrelativistic ones, while the lowest energy levels from the Dirac equation
describing Dirac and Majorana particles with the BC-MIT \cite{Chodos1,Chodos2} 
are lower than their nonrelativistic limits. The energy levels for a Majorana bouncer is identical to the ones for a Dirac bouncer. The transition frequency between two neighboring energy eigenstates for a Dirac or Majorana bouncer is greater than the counterpart for a KG bouncer, which is in turn greater than the one for a nonrelativistic bouncer. Nevertheless, the relativistic corrections to the transition frequencies at the lowest few energy levels are too small to be detected by current technology in experiments using either neutrons or Ps atoms as bouncers.

We further study the probability, normal probability current, and scalar densities of the above bouncing particles. The behaviors of probability densities for nonrelativistic and KG bouncers around the boundary are similar, both vanish around the boundary under the Dirichlet boundary conditions. For a positive-energy Dirac bouncer, the probability density does not vanish around the boundary, and on top of this nonzero, static probability density, there would be {\it Zitterbewegung} for a Majorana bouncer depending on its spin orientation. The same {\it Zitterbewegung} can also be observed in the probability current densities of Majorana bouncers, although in some special conditions it disappears and then the behavior of the Majorana bouncer does coincide with the behavior of a Dirac bouncer in the same conditions. In the nonrelativistic limit, anyway, these effects fade away. 
The spin-dependent {\it Zitterbewegung} here is not produced by the BC-MIT, but comes from the interference between the positive- and negative-energy components of the Majorana wave function.
Under the BC-MIT, the probability current and scalar densities for both Dirac and Majorana particles vanish around the boundary. 
In general, the probability current density of a positive-energy Dirac bouncer in bound states vanishes everywhere, while a Majorana bouncer does not. In contrast, the scalar density of a Majorana particle vanishes everywhere, while a Dirac particle does not.

For the Schr\"{o}dinger and KG equations, we introduced the Dirichlet boundary conditions that the wave functions vanish at the boundary \cite{Saa}.
The Dirichlet boundary condition for a Dirac or Majorana wave function $\psi$, however, will leads to a trivial solution with $\psi=0$ everywhere \cite{Boulanger}. Thus, we adopt the BC-MIT, which implies both the normal probability current and the scalar densities of the particle vanish at the boundary, while the wave function there can be nonzero. If one introduces a finite potential step for the floor instead of an infinite potential barrier of an ideal mirror \cite{Kawai99}, the wave function of a KG or nonrelativistic particle will get a finite penetration depth into the floor (cf. the probability densities of Dirac and Majorana bouncers in FIG. \ref{probabilitydensity}). This implies a lower expectation value of altitude and so a lower eigen-energy of the particle compared with those under the boundary conditions of ideal mirrors. With a finite potential step, even a KG particle may have the lowest energy levels lower than those of a nonrelativistic particle bouncing above an ideal mirror.

\section*{Acknowledgments}

We thank Y. Kojima, Y. Kamiya, J. Murata, T. Yoshioka, T. Wakasa, J. Soda, Y. Nambu, S. Kanno, N. Matsumoto, K. Ishikawa, M. Okawa, R. B. Mann, A. S. Adam, T.-H. Li, and B. L. Hu  for useful discussions and comments. This work was supported by Ministry of Education, Culture, Sports, Science and Technology (MEXT)/Japan Society for the Promotion of Science (JSPS) KAKENHI Grant No. 15H05895, No.~17K05444 (KY)/Sasakawa Scientific Research Grant from the Japan Science Society/JSPS KAKENHI Grant No. 20J22946 (KU). AR is supported by Japanese Government (Monbukagakusho: MEXT) Scholarship. 
SYL is supported by the Ministry of Science and Technology of Taiwan under Grant No. MOST 109-2112-M-018-002 and in part by the National Center for Theoretical Sciences, Taiwan.

\begin{appendix}

\section{Quantum mechanical analysis of relativistic corrections to energy levels}
\label{perturbation}

In what follows, we calculate the relativistic correction to the energy levels of a  
quantum mechanical particle in a perturbative approach (see e.g.,~\cite{Mann, Griffiths}) in Rindler coordinates (e.g.,~\cite{HIUY}).

\subsection{Hamiltonian for a classical particle in Rindler coordinates}

The action of a relativistic particle moving in curved spacetime is given by 
\begin{eqnarray}
S=-m\int \sqrt{g_{\mu\nu}\frac{dx^\mu}{d\lambda}\frac{d x^\nu}{d\lambda}}d\lambda,
\end{eqnarray}
where $\lambda$ is an affine parameter for the worldline of the particle.
With the line element (\ref{Rindlermetric}) of Rindler coordinates, 
we parametrize the worldline in $\eta$ and so the action reads
\begin{eqnarray}
S=-\int m\sqrt{e^{2a\xi}\left(1- \dot{\xi}^2 \right)}d\eta = \int L d\eta,
\end{eqnarray}
where $\dot{\xi}\equiv d\xi/d\eta$ and we have suppressed the transverse motion in the $x$ and $y$ directions.
The conjugate momentum of the position $\xi(\eta)$ of the particle is defined by 
\begin{eqnarray}
P=\frac{\delta S}{\delta \dot{\xi}}
=\frac{me^{2a\xi}\dot{\xi}}{\sqrt{e^{2a\xi}(1-\dot{\xi}^2)}},
\label{pxperp}
\end{eqnarray}
which implies $\dot{\xi} 
=\sqrt{P^2/(P^2+m^2e^{2a\xi})}$. Then the Hamiltonian is given by
\begin{eqnarray}
H=P\dot{\xi}-L=me^{a\xi}\sqrt{\frac{P^2}{m^2e^{2a\xi}}+1}.
\end{eqnarray}
Assuming the rest-mass energy dominates ($P^2/m^2 \ll 1$), 
the acceleration is small and the particle is not highly excited ($a\xi \ll 1$), we have
\begin{eqnarray}
H\approx H^{}_{\rm NR} \equiv m+\frac{P^2}{2m}+ma\xi+\frac{ma^2\xi^2}{2}-\frac{P^4}{8m^3}-\frac{a\xi P^2}{2m}
\label{NRlimit}
\end{eqnarray}
as $\sqrt{1+x}=1+\frac{1}{2}x-\frac{1}{8}x^2+O(x^3)$ and $e^{x}=1+x+\frac{1}{2}x^2+O(x^3)$ 
\footnote{
The nonrelativistic Hamiltonian (\ref{NRlimit}) is consistent with the lowest orders of the Foldy-Wouthuysen transformed Hamiltonian for Dirac particles \cite{FW} in Rindler coordinates (\ref{Rindlermetric}) in the limit $a^2 \ll m$. 
Our nonrelativistic Hamiltonian (\ref{NRlimit}) is, however, not exactly the same as those obtained in Refs. \cite{Hehl, Silenko, Boulanger}, where they applied the metric first given by Kottler \cite{Kottler14}, 
rather than our (\ref{Rindlermetric}) by Lass \cite{Lass63}. In this paper, we adopt Eq. (\ref{Rindlermetric}) since when ${\bm x}^{}_\perp$ are suppressed, ($\eta, \xi$) in (\ref{Rindlermetric}) are the radar coordinates, which are naturally measurable for a uniformly accelerated observer localized at $\xi=0$ \cite{Lin20a}.}
.
Since the rest-mass energy of a particle is irrelevant to its motion, we exclude the first term $m$ from the Hamiltonian
and redefine the subtracted Hamiltonian for the particle as
\begin{eqnarray}
H^{}_{\rm NR} \equiv H^0 + H',  
\end{eqnarray}
where 
\begin{equation}
  H^0={P^2}/{2m}+ma\xi
\end{equation}	
will be the nonrelativistic Hamiltonian if we replace $\xi$ here by $z$ in Section \ref{NRQM}, and
\begin{equation}
  H'=\frac{ma^2\xi^2}{2}-\frac{P^4}{8m^3}-\frac{a\xi P^2}{2m}
\end{equation}
is the relativistic corrections, which is a perturbation in our approximation.

\subsection{Relativistic corrections}

Let us treat $\xi$ in the above Hamiltonians as $z$ in Section \ref{NRQM} and quantize the system by introducing $[~\xi, P~] = i$.
As we have reviewed in Section \ref{NRQM}, the energy levels given by the nonrelativistic Hamiltonian are 
\begin{eqnarray}
E_n=\langle\psi_n|H^0|\psi_n\rangle=ma\mathcal{B} \zeta_n,
\end{eqnarray}
where $|\psi_n\rangle$ is the $n$-th energy eigenstate, whose wave function 
$\psi_n(z) = \langle z | \psi_n\rangle$ is given in (\ref{NRwavefunc}), 
and $-\zeta_n$ is the $n$-th zero of the Airy function.
Note that $\psi_n(z)$ here satisfies the Dirichlet boundary condition $\psi|_{z=0}=0$. 

The first-order correction to the energy level is given by
\begin{eqnarray}
\tilde{ \cal E}_n^{(1)} &\approx& \langle\psi_n|H^\prime|\psi_n\rangle \nonumber\\
&=&\frac{ma^2}{2}\langle \xi^2\rangle-\frac{1}{8m^3}\langle P^4\rangle-\frac{a}{2m}\langle \xi P^2\rangle,
\label{firstorderpt}
\end{eqnarray}
where all the expectation values are taken with respect to the state $|\psi_n\rangle$. 
There seems to be an ambiguity of ordering in the last term because $\xi$ and $P^2$ do not commute. 
Fortunately, $[\xi, P^2 ] = 2iP$ and $\langle \psi_n | P | \psi_n \rangle = 0$ in our case
($\int_0^\infty {\rm Ai}(\zeta-\zeta_n) {\rm Ai'}(\zeta-\zeta_n) d\zeta \propto {\rm Ai}^2(\zeta-\zeta_n)|_{\zeta=0}^\infty = 0$).
From the Schr\"odinger equation (\ref{timeindependentsch}), one has
\begin{eqnarray}
P^2|\psi_n\rangle=2m(E_n-ma\xi)|\psi_n\rangle,
\end{eqnarray}
so (\ref{firstorderpt}) can be simplified to
\begin{eqnarray}
  \tilde{\cal E}_n^{(1)}\approx -\frac{(E_n)^2}{2m}+ma^2\langle\xi^2\rangle.
\end{eqnarray}
where $E_n=ma\mathcal{B}\zeta_n$ refers to the energy level for a nonrelativistic particle. 
The first term $-(E_n)^2/(2m)$ gives $-ma^2\beta^2\zeta_n^2/2$, while
\begin{eqnarray}
ma^2\langle\xi^2\rangle &=& ma^2 \mathcal{B}\int_0^\infty \psi_n^*(\zeta) \xi^2 \psi_n(\zeta) d\zeta
=ma^2\mathcal{B}^3 {\cal N}_n^2 \int_0^\infty \zeta^2({\rm Ai}(\zeta-\zeta_n))^2 d\zeta
= m a^2 \mathcal{B}^2 \frac{8}{15} \zeta_n^2
\end{eqnarray} 
with $\zeta=\xi/\mathcal{B}$. Then we recover the correction to the energy levels given in (\ref{KGEnCorrection}).

Note that the above relativistic corrections are positive, as we expected for a KG particle in the presence of the perfect-mirror boundary condition at $\xi=0$.

\end{appendix}


\end{document}